\documentclass[12pt]{article}
\RequirePackage[OT1]{fontenc}
\RequirePackage{amsthm,amsmath}
\usepackage{breqn}
\usepackage{soul}
\usepackage[numbers,sort&compress]{natbib}
\RequirePackage[citecolor=blue,urlcolor=blue]{hyperref}
\usepackage{hyperref}
\usepackage{amsfonts}
\usepackage{amssymb}
\usepackage{graphicx}
\usepackage{setspace}

\numberwithin{equation}{section}
\newcommand{\bOm}{{\boldsymbol \Omega}}

\def\b1{{\mathbf 1}}
\def\b0{{\mathbf 0}}

\def\bA{{\boldsymbol A}}
\def\bB{{\boldsymbol B}}

\def\bk{{\boldsymbol k}}

\def\bT{{\boldsymbol T}}

\def\bp{{\boldsymbol p}}
\def\bP{{\boldsymbol P}}
\def\bQ{{\boldsymbol Q}}

\def\bx{{\boldsymbol x}}

\def\bT{{\boldsymbol T}}

\def\bs{{\boldsymbol s}}

\def\cE{{\mathcal E}}
\def\cK{{\mathcal K}}
\def\cH{{\mathcal H}}

\def\bbI{{\mathbb I}}

\graphicspath{%
    {converted_graphics/}
    {/}
}

\begin{document}

\title{Canonical equivalence of a particle in a magnetic field to a simple oscillator}
\author{Henryk Gzyl\\
Centro de Finanzas IESA, Caracas, Venezuela.\\
 henryk.gzyl@iesa.edu.ve}
\date{}
 \maketitle

\setlength{\textwidth}{4in}

\vskip 1 truecm
\baselineskip=1.5 \baselineskip \setlength{\textwidth}{6in}
\begin{abstract}
It is proved that a classical (respec. quantum) system consisting of a particle in a constant magnetic field is canonically (respec. unitarily) equivalent to a 2-dimensional harmonic oscillator plus a free particle. It is also shown that the eigenvectors of the discrete spectrum are entangled states of the 2-dimensional harmonic oscillator.
\end{abstract}

\noindent {\bf Keywords}: Canonical transformation, Particles in magnetic fields, Harmonic Oscillators   \\


\section{Introduction and Preliminaries} 
The Hamiltonian description of the motion of a charged particle moving in a static, homogeneous magnetic field is the springboard to the quantum mechanical description, and it is presented in many ways. See \cite{G}, \cite{BJ}, \cite{Bal} or \cite{St} as representative of a long list of references.

Here we consider a complementary approach, consisting of performing a canonical transformation that eliminates the magnetic field (or rather, its vector potential) from the Hamiltonian, and transforming it into a system consisting of a two-dimensional harmonic oscillator plus a free particle, or to be proper, of a particle under the influence of a planar restoring force and moving freely along the third coordinate.

When the transformation is implemented as a unitary transformation between the corresponding Hilbert spaces, we find that the Hamiltonian of the particle in the magnetic field has a spectrum consisting of a discrete part overlapped by a continuous part. And more importantly, we obtain a representation of the eigenstates of the discrete part as an entanglement of oscillator states with time-dependent coefficients.

the rest of the paper is organized as follows. In the next section, we recall the classical Hamiltonian approach. In order to bring forth the canonical equivalences we describe the vector potential $\bA$ using a matrix. We then establish the canonical equivalence between the particle in a magnetic field and a two-dimensional harmonic oscillator plus a free particle. In the second section, we consider the quantum version of the problem, and show how to implement the canonical transformation by a time-dependent, unitary, transformation. We verify the unitarity and find explicitly the wave functions of the particle in the magnetic field in terms of the wave functions of the harmonic oscillator and the free particle. Furthermore, we determine how to write the states of discrete energy (the Landau states) of the particle, can be written as entangled states of the two-dimensional oscillator.

\subsection{The Hamiltonian approach}
That the Hamiltonian function to describe a particle in a homogeneous magnetic field is given by (see \cite{G} or \cite{FLS}, for example):
\begin{equation}\label{H1}
H(\bx,\bp) = \frac{1}{2m}\bigg(\bp - \frac{q}{c}\bA(\bx)\bigg)^2.
\end{equation}
Here, $\bA(\bx)=\bB\times\bx/2,$ and it is convenient for our approach to rewrite de vector product $\bB\times\bx$ as $\bOm\bx$ where
\begin{equation}\label{ang}
\bOm = \begin{pmatrix}0 & -\omega_3 & \omega_2\\
                \omega_3 & 0 & -\omega_1\\
								-\omega_2 & \omega_1 & 0,
				\end{pmatrix} 
\end{equation}			
where we put $\omega_i=\frac{qB_i}{mc}.$ We simplify, and suppose that the magnetic field is oriented along the $z-$axis, i.e. $\bB=B\hat{\bk}$ and write $\bOm$ as

\begin{equation}\label{ang1}
\bOm = \begin{pmatrix}0 & -\Omega &0 \\
                \Omega & 0 & 0\\
								0 & 0 & 0
				\end{pmatrix} = \begin{pmatrix} \bOm_0 & \b0\\	
			\b0^t & 0\end{pmatrix}\;\;\;\mbox{with}\;\;\;\bOm_0=\begin{pmatrix} 0 & -\omega\\
				                                 \omega & 0\end{pmatrix}
\end{equation}			
The vector $\b0$ is two dimensional zero vector, and the superscript ``t'' stands for the transpose of the indicated object. Also, keep in mind that $\bOm_0^t\bOm_0=\omega^2\bbI.$ Below we make extensive use of the fact that
\begin{equation}\label{rot1}
U(t) = e^{t\bOm} = \begin{pmatrix}e^{t \bOm_0} & \b0\\	
			\b0^t & 1\end{pmatrix}\;\;\;\mbox{with}\;\;\;
			e^{t\bOm_0}=\begin{pmatrix} \cos(\omega t) & \sin(\omega t)\\
				                                \sin(\omega t) & \cos(\omega t)\end{pmatrix}
\end{equation}

\subsection{Canonical equivalence to the simple oscillator}
Since $U(t)$ denotes a rotation about the $\hat{\bk}-$axis, it is convenient to separate notationally the components in the plane from those along the $\hat{\bk}-$axis. For that,
put $\bx=(\bar{\bx},x_3)$ and  $\bp=(\bar{\bp},p_3)$ with $\bar{bx}=(x_1,x_2)$ and $\bar{\bp}=(p_1,p_2)$ respectively. With these notations, we can write the Hamiltonian \eqref{H1} as

\begin{equation}\label{H3}
\begin{aligned}
H(\bx,\bp)& = \frac{1}{2m}\bigg(\bp - \frac{m}{2}\bOm\bx\bigg)^2\\
               &=\frac{1}{2m}\langle\bar{\bp},\bar{\bp}\rangle -\frac{1}{2}\langle\bar{\bp},\bOm_0\bar{\bx}\rangle + \frac{m\omega^2}{2}\langle\bar{\bx},\bar{\bx}\rangle + \frac{1}{2m}p^2_3.
\end{aligned}
\end{equation}
Consider now the canonical transformation (see \cite{A} or \cite{G}) $(\bx,\bp)\to (\bQ,\bP),$ generated by:
\begin{equation}\label{gf}
F_2(\bx,\bP,t) = \langle\bx,U(-t/2)\bP\rangle=\langle\bar{\bx},e^{-t\bOm_2/2}\bar{\bP}\rangle +x_3P_3.
\end{equation}													
 The transformation equations (see \cite{A} or \cite{G}) are:
\begin{equation}\label{teq1}
\begin{aligned}
\bQ = \nabla_{\bP}F_2(\bx,\bP,t);\;\;\;&\bp = \nabla_{\bx}F_2(\bx,\bP,t)\\
K = H +& \frac{\partial F_2}{\partial t}
\end{aligned}
\end{equation}
The second equation is obtained using the first to replace the old variables with the new ones. The result is:
\begin{equation}\label{teq2}
\begin{aligned}
\bar{\bQ} &= e^{t\bOm_0/2}\bar{\bx};\;\;\bar{\bP}=e^{t\bOm_0/2}\bar{\bp}, Q_3=z_3, P_3=p_3.\\
K(\bQ,\bP)& = \frac{1}{2}\langle\bar{\bP},\bar{\bP}\rangle+ \frac{m\omega^2}{2}\langle\bar{\bQ},\bar{\bQ}\rangle + \frac{1}{2}P_3^2.
\end{aligned}
\end{equation}			
We used the fact that 	$\langle\bar{\bP},\bar{\bP}\rangle=\langle\bar{\bp},\bar{\bp}\rangle,$ $\langle\bar{\bQ},\bar{\bQ}\rangle=\langle\bar{\bx},\bar{\bx}\rangle,$ and that 
$$\partial F_2/\partial t =\frac{1}{2} \langle\bar{\bp},\bOm_0\bar{\bx}\rangle,$$
using \eqref{teq2} after differentiating. 
In the new coordinates, we have a two-dimensional harmonic oscillator plus a free motion along the $z-$axis.

\section{The quantum particle in a homogeneous magnetic field}
Let us write $\cK$ and $\cH$ for the state spaces in which the time evolution is described by

\begin{gather}
\hat{K} =-\frac{\hbar^2}{2m}\Delta_Q - \frac{\hbar}{2m}\frac{\partial^2}{\partial Q^2_3}  + \frac{m\omega^2}{2}\langle\bar{\bQ},\bar{\bQ}\rangle.  \label{ham1}\\
\hat{H} =-\frac{\hbar^2}{2m}\Delta_x - \frac{\hbar^2}{2m}\frac{\partial^2}{\partial x^2_3}  -\frac{i\hbar}{2}\langle\bOm\bar{\bx},\nabla_x\rangle +  \frac{m\omega^2}{2}\langle\bar{\bx},\bar{\bx}\rangle. \label{ham2}
\end{gather}
We used the fact that when acting on functions of $\bx, $ we have:
$$\langle\bOm\bx,\nabla_{\bx}\rangle + \langle \nabla_{\bx},\bOm\bx,\rangle = 2\langle\bOm\bx,\nabla_{\bx}\rangle + div(\bOm\bx) = 2\langle\bOm\bx,\nabla_{\bx}\rangle$$
because $div(\bOm\bx)=tr(\bOm)=0.$

The operators $\hat{K},$  respectively $\hat{H},$ act on generic square-integrable functions denoted by $\Phi(\bQ,t)$ and, respectively, $\Psi(\bx,t),$ in the coordinate representation of the quantum states. Let us now see that the canonical transformation \eqref{teq1} generated by \eqref{gf} can be represented by a time-dependent unitary transformation that maps states $\Phi(\bQ,t)$ evolving according to $\hat{K}$ onto states $\Psi(\bx,t)$ evolving according to $\hat{H}.$
We define $\bT_t\Phi(\bQ,t)$ as follows: First write
\begin{equation}\label{momrep}
\tilde{\Phi}(\bP,t) = \frac{1}{(2\pi)^{3/2}}\int\Phi(\bQ,t)e^{-i\langle\bP,\bQ\rangle/\hbar}d\bQ.
\end{equation}
Now, we exploit the linearity of $F_2$ and define the unitary representation of the transformation generated by eqref{gf} as follows:
\begin{equation}\label{repct}
\Psi(\bx,t) = \bT_t\Phi(\bQ,t) = \frac{1}{(2\pi)^{3/2}}\int\tilde{\Phi}(\bP,t)e^{iF_2(\bx,\bP,t)}d\bP =
\Phi(U(t/2)^{\dag}\bx,t),
\end{equation}
Since $\Omega$ is anti-symmetric $U^{\dag}(t)=U(-t).$ Since at $t=0,$ $F_2(\bx,\bP,0)$ generates the identity transformation, applying $\bT_0,$ one recovers the coordinate representation from the momentum representation. Also, if $\Phi_1$ and $\Phi_2$ are two states in $\bQ-$representation, and $\Phi_1,$ $\Psi_2$ are as in \eqref{repct},  using the integral representation of the $\delta-$function and the invariance of $d\bQ$ under rotations, we obtain:

\begin{equation}\label{unty}
\begin {aligned}
&\langle\Psi_1(t),\Psi_2(t)\rangle \\
&  = \int\int\delta\big(U(-t/2)\bQ_1 - U(-t/2)\bQ_2\big)\bar{\Phi}_1(U(-t/2)\bQ_1,t)\Phi_1(U(-t/2)\bQ_2,t)d\bQ_1d\bQ_2\\
& = \langle\Phi_1(t),\Phi_2(t)\rangle.
\end{aligned}
\end{equation}
That is, the transformation \eqref{repct} is unitary. Since in the momentum representation $\hat{\bP}\tilde{\Phi}(\bP,t)=\bP\tilde{\Phi}(\bP,t)$, from \eqref{repct} it follows that
$$-i\hbar\nabla_{\bx}\big( \bT_t\Phi(\bQ,t)\big)=U(t/2)\frac{1}{(2\pi)^{3/2}}\int\big(\bP\Phi(\bP,t)\big)e^{iF_2(\bx,\bP,t)}.d\bP$$

That is the momentum operator in the $\bQ-$coordinates transforms into the momentum operator in the $\bx$ coordinates by the action of $\bT_t.$ This is just a different way of using the chain rule and \eqref{teq2} to obtain $\partial x_i=U_{i,j}(t/2)\partial Q_j.$ The same holds for the coordinate operators. 

It takes but a short computation to see that if $\Psi$ and $\Phi$ are related by \eqref{repct}, then:
\begin{equation}\label{consit}
i\hbar\frac{d}{dt}\Phi(\bQ,t) = \hat{K}\Phi(\bQ,t)\;\;\;\Longrightarrow\;\;\;i\hbar\frac{d}{dt}\Psi(\bx,t) = \hat{H}\Psi(\bx,t)
\end{equation}
From this and from \eqref{unty} it follows that eigenstates of $\hat{K},$ are mapped onto eigenstates of $\bar{H}.$  The spectrum of $\bar{K}$ is an overlap of a discrete part, namely the spectrum of a two-dimensional harmonic oscillator, plus a continuous part, namely the spectrum of a one-dimensional free particle. We have in a self-explanatory notation
\begin{equation}\label{spec}
E(n_1,n_2,k) = \hbar(n_1+\frac{1}{2}) +  \hbar(n_2+\frac{1}{2}) +\frac{\hbar^2k^2}{2m},
\end{equation}
where $n_1,n_2$ are positive integers and $k$ is a real number. The corresponding eigenstates are entangled eigenstates of the two-dimensional harmonic oscillator, where the entanglement is caused by the rotation in the plane perpendicular to the direction of the magnetic field, with angular speed half of the cyclotron frequency $\omega=qB/mc.$ It is clear that it suffices to examine the eigenvalues of the discrete part. For that recall that the eigenvectors of the one-dimensional harmonic oscillator are:

\begin{equation}\label{evec1}
\psi_n(x) = \bigg(\frac{\alpha}{\pi^{1/2}2^nn!}\bigg)^{1/2}H_n(\alpha x)e^{-\frac{1}{2}x^2},
\end{equation}
Where $H_n(x)$ is the Hermitte polynomial of degree $n,$ and $\alpha=\big(m^{3/2}\omega/\hbar\big)^{1/2}.$ What is important for us at this point, is that the Hermite polynomials are obtained from their generating function as follows:

\begin{equation}\label{genfun}
\exp\big(-s^2 + 2sz\big) = \sum_{n=0}^\infty \frac{s^n}{n!}H_n(z).
\end{equation}

From \eqref{repct} if follows that the transform of the eigenvector $\psi_n(Q_1)\psi_m(Q_2)$ is
$\psi_n\big((U(-t/2)\bx)_1\big)\psi_m\big((U(-t/2)\bx)_2\big),$ where, for example, $(U(-t/2)\bx)_1$ denotes the first component of the vector $U(-t/2)\bx,$ to wit, $U_{1,1}x_1+U_{1,2}x_2,$ after dropping reference to the time variable. In order to expand $\psi_n\big((U(-t/2)\bx)_1\big)\psi_m\big((U(-t/2)\bx)_2\big)$ in terms of  $\psi_k(x_1)\psi_l(x_2)$ we note that, according to \eqref{evec1}, and since:
$$(U(-t/2)\bx)^2_1 + (U(-t/2)\bx)^2_2 = \|U(-t/2)\bar{\bx}\|^2 = \|\bar{\bx}\|^2 = x^2_1+x^2_2$$
it suffices to concentrate on the corresponding Hermite polynomials. For that consider: 
$$
\sum_{n,m=0}^\infty \frac{s_1^ns_2^m}{n!m!}H_n((U(-t/2)\bx)_1)H_m((U(-t/2)\bx)_2) = e^{-
s_2^2-s_2^2 +2\big(s_1(U(-t/2)\bx)_1+s_2(U(-t/2)\bx)_2\big)}
$$
If we put $\bs=(s_1,s_2,0)^t,$ it is clear that the exponent in the right-hand side can be written (again, dropping the argument of U)  as:
$$\langle\bs,\bs\rangle = \langle U\bs,U\bs\rangle,\;\;\;\mbox{and}\;\;\;\langle\bs,U\bx\rangle =\langle U^t\bs,\bx\rangle.$$
Now, invoke \eqref{genfun}, to obtain

\begin{equation}\nonumber
\begin{aligned}
&\sum_{p,q=0}^\infty \frac{\big((U^t\bs)_1\big)^p}{p!}\frac{\big((U^t\bs)_2\big)^q}{q!}H_p(x_1)H_q(x_2)\\
& = \sum_{p,q=0}^\infty \frac{\big((U^t_{1,1}s_1+U^t_{1,2}s_2\big)^p}{p!}\frac{\big((U^t_{2,1}s_1+U^t_{2,2}s_2\big)\big)^q}{q!}H_p(x_1)H_q(x_2) \\
& \;\;\;\;\;\;\mbox{\rm{using the binomial expansion and collecting terms}}\\
&=\sum_{p,q=0}^\infty\sum_{k_1=0}^p\sum_{k_2=0}^q \frac{\big((U^t_{1,1})^{k_1}s_1^{k_1}(U^t_{1,2})^{p-k_1}s^{p-k_1}_2\big)}{k_1!(p-k_1)!}\frac{\big((U^t_{2,1})^{k_2}s_1^{k_2}(U^t_{2,2})^{q-k_1}s^{q-k_1}_2\big)\big)^q}{k_2!(q-k_2)!}H_p(x_1)H_q(x_2) \\
&\;\;\;\;\;\;\mbox{\rm{Exchanging the order of summations}}\\
&=\sum_{k_1,k_2=0}^\infty\frac{(U^t_{1,1})^{k_1}s_1^{k_1}U^t_{2,1})^{k_2}s_1^{k_2}}{k_1!k_2!}\sum_{p=k_1}^\infty\sum_{q=k_2}^\infty\frac{(U^t_{1,2})^{p-k_1}s^{p-k_1}_2(U^t_{2,2})^{q-k_1}s^{q-k_1}_2}{(p-k_1)!(q-k_1)!}H_p(x_1)H_q(x_2)\\
&\;\;\;\;\;\;\mbox{Shifting the starting point of the inner summations}\\
&=\sum_{k_1,k_2=0}^\infty\frac{(U^t_{1,1})^{k_1}s_1^{k_1}U^t_{2,1})^{k_2}s_1^{k_2}}{k_1!k_2!}\sum_{p,q=0}^\infty\frac{(U^t_{1,2})^ps^p_2(U^t_{2,2})^qs^q_2}{p!q!}H_{n+k_1}(x_1)H_{m+k_2}(x_2)\\
&\;\;\;\;\;\;\;\;\mbox{Rearranging the order of summation and collecting terms}\\
&=\sum_{k_1,k_2=0}^\infty\sum_{p,q=0}^\infty\frac{(U^t_{1,1})^{k_1}U^t_{2,1})^{k_2}(U^t_{1,2})^p(U^t_{2,2})^q}{k_1!k_2!p!q}s^{k_1+p}_1s^{k_2+q}_2H_{n+k_1}(x_1)H_{m+k_2}(x_2)\\
&\;\;\;\;\;\;\;\;\mbox{After a change of summation variables and collecting powers with the same basis}\\ 
&=\sum_{n_1,n_2=0}^\infty\sum_{l_1=0}^{n_1}\sum_{l_2=0}^{n_2}\frac{C(n_1,n_2,l_1,l_2,t)}{n_1!,n_2!}{n_1\atopwithdelims()l_1}{n_2\atopwithdelims()l_2}s^{n_1}_1s^{n_2}_2H_{l_1+l_2}(x_1)H_{(n_1+n_2)-(l_1+l_2)}(x_2).
\end{aligned}
\end{equation}
 As $U_{1,1}=U_{2,2}$ and $U_{1,2}=U_{2,1}$ are, respectively, $\cos(\omega t)$ and $\sin(\omega t),$ the coefficient $C(n_1,n_2,l_1,l_2,t)$ is a product of trigonometric functions. Explicilty:
 
 \begin{equation}\label{coef1}
 C(n_1,n_2,k,l,t) = \bigg(\cos(\omega t)\bigg)^{(n_1+l_1)}\bigg(\sin(\omega t)\bigg)^{(n_2+l_2)}.
 \end{equation} 
 
 Now, invoke \eqref{evec1} and \eqref{genfun}, Modify the coefficients as explained below, and obtain:
\begin{equation}\label{entang}
\begin{aligned}
\psi_n&\big((U(-t/2)\bx)_1\big)\psi_m\big((U(-t/2)\bx)_2\big)\\
& =\sum_{l_1=0}^{n_1}\sum_{l_2=0}^{n_2}D(n_1,n_2,l_1,l_2,t){n_1\atopwithdelims()l_1}{n_2\atopwithdelims()l_2}s^{n_1}_1s^{n_2}_2\psi_{l_1+l_2}(x_1)\psi_{(n_1+n_2)-(l_1+l_2)}(x_2).
\end{aligned}
\end{equation}
We put 
\begin{equation}\label{coef2}
D(n_1,n_2,l_1,l_2,t) = C(n_1,n_2,l_1,l_2,t)\bigg(\frac{\big(l_1+l_2\big)!\big((n_1+n_2)-(l_1+l_2)\big)!}{n_1!n_2!}\bigg)^{1/2}.
\end{equation}
Thus the eigenvectors of the corresponding to the discrete part of the spectrum, having energy  $\cE_{n_1+n_2}=\hbar(n_1+\frac{1}{2}) +  \hbar(n_2+\frac{1}{2})$ are an entanglement of $1-$dimensional oscillator states whose energy adds up to $\cE_{n_1+n_2}.$  For different approaches to this problem, see \cite{Bal}, \cite{BJ}, and a totally different approach, see \cite{D}.

{\bf Declaration of competing interests} We have no competing interests to declare.
\end{document}